\documentclass[preprintnumbers,amsmath,amssymb,floatfix,superscriptaddress,twocolumn]{revtex4}
\usepackage{epsfig}
\usepackage{graphicx} % Include figure files
\usepackage{bm}       % bold math
\usepackage{color}
\usepackage{epsfig}
\usepackage{subfigure}
\usepackage{epstopdf}

\begin{document}
\title{Classification of macroscopic quantum effects}

\author{Tristan Farrow}
\email{t.farrow1@physics.ox.ac.uk}
\affiliation{Department of Physics, University of Oxford, Parks Road, Oxford, OX1 3PU, UK}
\affiliation{Centre for Quantum Technologies, National University of Singapore, 3 Science Drive 2, Singapore 117543}
\author{Vlatko Vedral}
\email{v.vedral1@physics.ox.ac.uk}
\affiliation{Department of Physics, University of Oxford, Parks Road, Oxford, OX1 3PU, UK}
\affiliation{Centre for Quantum Technologies, National University of Singapore, 3 Science Drive 2, Singapore 117543}

%\date{\today}

\begin{abstract}
We review canonical experiments on systems that have pushed the boundary between the quantum and classical worlds towards much larger scales, and discuss their unique features that enable quantum coherence to survive. Because the types of systems differ so widely, we use a case by case approach to identifying the different parameters and criteria that capture their behaviour in a quantum mechanical framework. We find it helpful to categorise systems into three broad classes defined by mass, spatio-temporal coherence, and number of particles. The classes are not mutually exclusive and in fact the properties of some systems fit into several classes. We discuss experiments by turn, starting with interference of massive objects like macromolecules and micro-mechanical resonators, followed by self-interference of single particles in complex molecules, before examining the striking advances made with superconducting qubits. Finally, we propose a theoretical basis for quantifying the macroscopic features of a system to lay the ground for a more systematic comparison of the quantum properties in disparate systems.


\end{abstract}

\maketitle

\section{Introduction}
One of the most fascinating questions in quantum physics is whether large objects, say cats, can show features of the strange quantum behaviour of atoms and particles. When Erwin Schr\"{o}dinger thought-up his Gedanken experiment in 1935 about a cat in a quantum superposition of states, so that it is dead and alive at the same time, he wanted to highlight the seeming contradictions -- not to say the absurdity -- of apprehending large objects through the framework of quantum mechanics.

Because Planck's constant is so small, quantum effects become imperceptible as objects grow in mass and complexity. Yet, there is nothing stopping us, in theory, from designing experiments where massive objects behave as if they were atoms in at least one degree of freedom. The boundary where quantum effects stop and classical physics takes over is blurry. Determining where that boundary lies is one of the most fascinating questions in physics and excites ongoing interest~\cite{leg80, vlatko08, vlatko04, vlatko06, shimizu, hwang, nim13, sang}. The challenge is largely technological, and as we will see, astounding experiments continue to push the quantum limit into the realm of macroscopic objects previously reserved for classical treatment.

It is extremely difficult to isolate massive objects from the environment as they constantly interact by exchanging photons, which leads to heating and therefore decoherence (i.e. the disruption of the constant phase relationship required for quantum states). The experimenter's challenge, then, is to find a macroscopic degree of freedom whose energy levels are separated by more than $k_B T$, where $k_B$ is Boltzmann's constant and $T$ is the temperature. To maintain the separation, the system is usually cooled to cryogenic temperatures that ensure the quantum state survives long enough for a measurement to be made.

We can identify at least two criteria that a macroscopic quantum state should meet: first, the state must be entangled and this entanglement must be verifiable experimentally, and second, it should be {\it macroscopically distinguishable}~\cite{peres, sang}, i.e. it must have macroscopic observables that we can use to discriminate different states that are combined into the overall entangled states. Although a systematic way of comparing the quantum features of disparate systems is currently lacking, we review and make some suggestions as to how this can be tackled in the section ``Discussion and conclusions''.

Let us now explore some recent experiments that have pushed the quantum limit to ever larger scales. They cover a wide variety of systems and cannot be easily described within a single framework. This imposes a case by case approach because no two systems can directly be compared when they are quantised in different degrees of freedom and use a wide variety of different macroscopic metrics. This accounts for the current lack of a clear measure of macroscopic quantumness.

\section{Massive objects}
Here, we explore the quantum properties of macroscopic bodies with a comparatively high fixed centre of mass. We look at interference experiments where macromolecules undergo diffraction at a grating, and then turn our attention to micromechanical resonators approaching the size of a human hair that operate in the quantum regime.

\subsection{Molecular interference}
Sending molecules one at a time at two slits can produce an interference pattern on a screen positioned beyond the slits. This signature of wave behaviour underpins de~Broglie's theory on the joint wave and particle character and propagation of massive objects. Since its inception in 1923, many experiments have successfully recovered the interference pattern due to objects at ever higher masses, starting with electrons, neutrons, atoms, dimers, and nowadays macromolecules. Some of the largest molecules to have been interfered are $C_{60}$ buckminster fullerenes (football-shaped carbon lattices) called buckyballs.

In the experimental set-up~\cite{zeilc601, zeilc602}, a hot molecular beam of $C_{60}$ molecules is produced by sublimation from an oven. The beam passes through rotating choppers that select the velocity of the molecules before they undergo collimation. The buckyballs finally impinge on a diffraction grating with $55~nm$-wide slits and periodicity $100~nm$. Detection of the interference pattern takes place not on a traditional screen, but by ionisation detection: molecules are ionised with a laser and detected in a vacuum chamber mounted on a translational scanning stage. Eventually, an interference pattern builds up showing a characteristic central peak and up to three higher-order peaks on both sides of the central maximum (limited by spectral coherence due to fluctuations in the velocity of the molecules).

The de~Broglie wavelength, $\lambda_B$, associated with a massive object is $\lambda_B=h/p$, where $h$ is Planck's constant and $p$ is the momentum. In the case of $C_{60}$ under experimental conditions,  $\lambda_B(C_{60}) \approx 3~pm$, which is more than 300 times smaller than the diameter of the buckyball ($\approx 10^{-9}~m$)~\cite{zeilc602}, and more than 50 times smaller than the slit width. Single molecules enter the grating one at a time (given a low flux) such that two separate molecules can never interfere. That an interference pattern can build-up under these conditions is deeply surprising because we are used to thinking of particles and molecules as point-like objects. In the quantum physical picture, however, they are treated as a wave during time-of-flight -- which we can in turn think of as a superposition of position states -- becoming point-like again at detection. Another quantum feature is that the position at arrival of individual incident molecules is entirely random and unpredictable.

This raises a fascinating question. Let us imagine that our senses were so sophisticated that we could resolve distances on the Planck scale. Would we then perceive objects, especially macroscopic ones, as behaving quantum mechanically?

Bohr's complementarity principle tells us that knowing which slit a particle enters destroys the interference pattern. During time-of-flight, hot molecules can emit thermal photons from the hundreds of mechanical degrees of freedom in their structure. They can give away potential information about their path and the slit they enter. But, for this to happen, the wavelength of the photons must be short enough to resolve the separation between neighbouring slits. So far, this has not been the case in experiments with $C_{60}$ due to the long wavelength attributed to thermal photons~\cite{zeilc602}. We cannot exclude the possibility that heavier and more complex molecules could leak useful information, which would kill the contrast in the interference pattern. The question is at what mass does this happen? Efforts are currently underway to interfere large proteins with order magnitude heavier masses than $C_{60}$, and which require more sophisticated interferometers~\cite{joe14}.

So far our only constraint has been that the molecule should not give access to which-path information as it grows in size and complexity.
Another constraint is the sophistication of laboratory equipment. Because massive objects have very short de~Broglie wavelengths, diffraction gratings must be fabricated to stringent parameters ranging in the tens of nanometres, which poses a significant technical challenge. A historical perspective allows us to be optimistic that advances in interferometry and detection technologies will further extend the quantum limit to larger bodies.

This is a natural point to ask if we can set an objective limit on the size of a body beyond which quantum superpositions collapse into classical mixtures. In 1964, Peres and Rosen~\cite{peres} took an operational view of the problem by setting an upper bound on the time it takes interference fringes to form when a massive body impinges on two slits. It can easily be shown (using the Fraunhofer limit of diffraction and the de Broglie wavelength of a massive object of momentum $mv$) that an approximate value of the time, $t$, that it takes to build up an interference pattern on a screen, is given by
\begin{equation}
t \approx \rho a^4 d / h
\end{equation}
where $\rho$ is the density of the body, $d$ is the distance between interference fringes, $h$ is Planck's constant, and $a$ is the size of the body (comparable to the separation between two slits). Assuming that $t < 10^{18}~s$ (the age of the universe) and $\rho \approx 1~g~cm^{-3}$ (a universal constant), we can estimate that $a_{max}$ and $m_{max}$ (the maximum size and mass of the body) are: $a_{max} < 1~cm$ and $m_{max} < 1~g$ .  Hence, objects whose mass and size exceeds those bounds cannot show quantum interference in a double slit experiment. Below this bound, and in different contexts, setting the boundary between the quantum and classical pictures may remain subjective and limited by other practical considerations.

\subsection{Micromechanical resonators}
Remarkable breakthroughs have been made in the field of micro-mechanical resonators~\cite{cleland10, aspel11, cleland14} over the last five years. These devices are currently enabling the study of quantum mechanics on scales of tens of microns almost visible to the naked eye.

Nano- and microresonators are heavy objects that tend to oscillate with relatively low frequencies (hence energies). To resolve their vibrational ground state, the following condition must be met: $k_B T < E_{vib}$, where $E_{vib}$ is the vibrational energy and $k_B$ is Boltzmann's constant. Hence extremely low temperatures are needed to decouple the slow-moving mechanical modes from phonons in the environment.  This challenging milestone eluded experimentalists for a long time until laser cooling offered a viable way of freezing out thermal phonons~\cite{aspel11, aspel10}.

More recently, an alternative approach was presented~\cite{cleland10} using a piezoelectric resonator whose the centre of mass is at rest, but which undergoes periodic volume increases. The resonator consists of a suspended piezoelectric blade with length around $\approx 40~\mu m$ and thickness $840~nm$. Since the volume-changes in the blade attain extremely high frequencies (gigahertz) ~\cite{cleland10}, the resonator needs less cooling than its slow-moving counterparts before its ground state can be resolved (due to a larger spacing in its energy levels at higher frequencies). In the actual experiment, the device was cooled to $25~mK$  in a dilution fridge, at which point it was shown to occupy the ground state with only 0.07 phonons populating the mechanical mode.

That same experiment achieved a striking second milestone by successfully demonstrating strong coupling between the mechanical mode of the resonator and a phase qubit in a superconducting circuit~\cite{cleland10, qubit}. Coherent transfer of energy between the resonator and the superconducting circuit was illustrated by measuring the phonon population in the mechanical mode, which oscillated with a characteristic Rabi frequency $\approx 130~MHz$. And to show that a state in a quantum superposition could also be coherently swapped between the two quantum systems, the superconducting qubit was prepared in a superposition of ground and excited states, $|g\rangle$ and $|e\rangle$, respectively, which was then transferred to the mechanical mode. The resulting state in the resonator was thus a superposition of the ground and excited vibrational modes, $|n=0\rangle$ and $|n=1\rangle$, which resulted under evolution in the entangled state, $|g1\rangle + |e0\rangle$, between the resonator and the quantum electrical circuit.

A fascinating experiment that still remains to be done is one where the resonator exists in a superposition of position states, so that it is in two places at once~\cite{aspel10}. This poses a significant challenge because such an experiment would have to use a resonator whose centre of mass undergoes periodic displacements that are large enough to be measured (unlike those of the piezoelectric resonator). Hence, it would run at a lower frequency and require extremely low temperatures -- below those accessible in a standard dilution fridge -- to resolve its ground state.

\section{Single particle in complex molecules}

Advances in spectroscopic techniques, like microphotoluminescence imaging and 2D spectroscopy, have enabled the observation of the coherent behaviour of individual quantum particles in a wide variety of ordered and disordered systems.

A fascinating example comes from photosynthesis. The protein responsible for harvesting light in photosynthetic bacteria, known as the Fenna-Matthews-Olson complex, has been shown to maintain long-lived spatio-temporal coherence of a photo-excited exciton~\cite{flem} even at room temperature~\cite{collini}. The implication is that the wavefunction of the exciton is in a coherent superposition over different sites throughout the molecule, so that it optimises the time it takes to propagate towards the protein's chemical reaction centre. The protein is approximately $8~nm$ in diameter and the particle's spatial delocalisation covers a small region within the molecule. Experiments~\cite{flem, collini} have shown that coherence persists for the entire duration of the energy transfer (on femtosecond timescales), which accounts for the fact that the efficiency of the energy transfer is almost perfect. It is very suprising that quantum coherence survives for so long in as highly disordered a system that is as 'hot and wet' as a protein. The result suggests that we could do even better in artifical systems that are engineered to be highly symmetric.

In a striking demonstration of the spatial coherence of a quantum particle on very large length scales, the centre of mass of a single exciton (a bound electron-hole quasi-particle) was shown to be delocalised along the entire length ($\approx 10~\mu m$) of an organic polymer chain~\cite{dubin06}. The polymer, a polydiacetylene chain decorated with urethane pending groups, behaved as a quantum wire characterised by long-range order and a highly symmetric lengthwise confining potential (with only local fluctuations due to build-up of elastic stress).

The experimental demonstration entailed an elegant adaptation of Young's double-slit experiment. Fluorescence due to the radiative recombination of a single delocalised exciton was collected through two slits separated by 1~${\mu m}$ (both slits selecting emission from two distinct points along the chain) and interfered. Interference peaks were generated in the detection apparatus to confirm the coherence of the light source (i.e. the single exciton). The contrast between interference fringes remained constant ($\approx 75\% $) along the entire length of the polymer, over 10~${\mu m}$, confirming that the exciton's centre of mass was delocalised over the entire distance, limited only by the natural length of the chain. Conventional ballistic and diffusive transport mechanisms cannot account for the delocalisation of the particle over tens of microns on the short timescales involved.

\section{Large numbers of particles}

A different class of large-scale quantum phenomena is that which involves the collective behaviour of many particles. In contrast to microscopic quantum states, which involve a low number of particles or degrees of freedom (e.g.~\cite{ian11, wein}), macroscopic states entail the coherent control and manipulation of some quantum degree of freedom of a large number of particles can claim to be macroscopic. Examples include Schr\"{o}dinger cat states~\cite{wine96}, high temperature superconductivity~\cite{vlatko04}, and angular momentum of photons~\cite{zeil12}) among others. Many-body physics is also an important area of study in condensed matter physics that investigates various exotic quantum effects in systems with a large number of particles \cite{Son09, Son12}.

It is worth noting an absence of consensus on how many particles should compose such a system before it is called macroscopic. Perhaps a more pragmatic approach to the problem is to consider the value of such a system as a resource, rather than purely in terms of its size or constituent number of particles (see the section ``Discussion and Conclusions").

Examples of collective quantum systems with relatively small numbers of entangled particles include ion traps~\cite{blatt08} and ultracold atoms in optical lattices~\cite{bloch08}. Bose-Einstein condensates (BEC), where thousands of atoms can exist in a cat-state  --- albeit the atoms sit too far apart to interact --- are also good examples. Their de~Broglie wavelength can extend over macroscopic distances (hundreds of microns), and, by analogy to Young's double-slit experiment, two BECs can undergo quantum interference across a double-well potential (acting as if it were a double-slit)~\cite{ket}.

Superconductors, which can be thought of as Bose-Einstein condensates made of coupled electron-pairs in the ground-state (Cooper pairs), represent the canonical example of a system showing large-scale quantum behaviour~\cite{clarke08}.

Cooper pairs participating in a supercurrent can be represented by a single macroscopic wavefunction rather than by individual wavefunctions for each carrier. Consequently, Cooper pairs can tunnel through a Josephson junction -- a nanometer-thick insulating layer sandwiched by two superconductors -- without any loss of phase throughout the entire condensate across the junction. And, if the supercurrent circulates around a circuit formed by a closed loop, the magnetic flux through the loop is quantised. A standard quantum mechanical operator representation is given by
\begin{equation}
[\delta, Q] = i2e
\end{equation}
where $\delta$ is the phase operator associated with the quantised phase difference between two superconductors on either side of a Josephson junction, and $Q$ is the charge operator associated with junction capacitance (related to the number-difference of Cooper pairs across the junction)~\cite{clarke08}. The non-commuting relationship between $\delta$ and $Q$ defines two regimes, where either the superconducting phase is well defined and charge is fuzzy, or only charge is well defined and phase is unspecified. This gives rise to quantisation in different degrees of freedom (phase, charge, and also flux), which different architerctures of superconducting circuits seek to optimise. Their striking property is that the specified degree of freedom can be macroscopic, i.e. a single wavefunction fully defines the state of the condensate throughout the macroscopic circuit. Let us now briefly consider the three main types of superconducting circuit architectures that give rise to macroscopic charge, phase, and flux qubits.

Charge qubits~\cite{nak} can be created on a micron-sized island of superconducting material coupled to a larger superconducting reservoir via a Josephson junction. Coupling is controlled by a gate bias that can be tuned so that a well-defined number of Cooper pairs (with fuzzy phase) tunnel one at a time from the reservoir into the island (called a Cooper pair box). The build-up of charges can result in a macroscopic quantum superposition of pair states, $|n \rangle$ and $|n + 1\rangle$, where $n$ is an integer number of charges. Charge qubits can comprise large ensembles of Cooper pairs with coherence times that can be extremely long (microseconds~\cite{paik}) when the state is shielded from the environment.

Phase qubits can be viewed as the conjugate states of charge qubits. They are formed by a single Josephson junction biased with an external current~\cite{martinisprl} whose energy profile follows an anharmonic washboard potential with a decreasing spacing between successive energy levels. This has earned the circuit the evocative title of ``macroscopic nucleus with wires''~\cite{clarke88}).
The particularity of phase qubits is that the condensates (involving a large number of Cooper pairs) all share the same macroscopic phase difference across the junction.

Flux qubits are a striking third example of macroscopic quantum states in superconductors. They are formed when a supercurrent circulates around a closed loop interrupted by a series of Josephson junctions~\cite{wal}, which gives rise to a superposition of clockwise and anti-clockwise currents in the loop, represented in the flux basis as
\begin{equation}
|\Phi \rangle = a| \uparrow \rangle + b| \downarrow \rangle
\end{equation}
where $| \Phi \rangle$ is the wavefunction of the macroscopic flux qubit consisting of quanta of magnetic flux pointing up, $| \uparrow \rangle$, and down, $| \downarrow \rangle$, with probability amplitudes $a$ and $b$ respectively.
Because the state can involve very large number of Cooper pairs ($\approx 10^9$), flux qubits are often described as macroscopic states.

So far, we have established that macroscopic quantum states in superconductors are specified by a single macroscopic parameter, rather than by individual ones for different particles. Let us now turn to the question as to whether we should worry about specifying a minimum number of particles participating in a superposition deemed to be macroscopic. After all, even a small number of Cooper pairs can give rise to macroscopic observables with a large magnitude.

Due to the fermion statistics and indistinguishability of Cooper pairs, it is easy to overestimate the number of particles in a state by many orders of magnitude. For flux qubits, say, we can estimate the difference in Cooper pairs flowing clockwise and counter-clockwise from the value of the superconducting current, $I_s$. This can in turn be written as $2e \Delta N/\Delta t \approx 2e \Delta N v_F/l$, where $l$ is the length of the superconductor (typically hundreds of $\mu m$) and $v_F$, the Fermi velocity ($\approx 10^6~ms^{-1}$).  Experimentally measured supercurrents can be on the order of $\mu A$, which leads to estimate of $\Delta N$ on the order of thousands (out of billions of potentially conducting electrons). This leads the authors of~\cite{whal10} to re-classify flux qubits as mesoscopic at best, rather than macroscopic.

That said, there is no fundamental upper bound on the size of superconducting qubits. With optimised circuit architectures and sophisticated fabrication facilities, we can be optimistic that superconducting qubits will attain surprising macroscopic dimensions, driven by the need for  coherent quantum-state control~\cite{leg02, vion, yu} in quantum information processing.

\section{Discussion and Conclusions}

When quantum macroscopicity is analysed theoretically, the states that undisputedly arise as prime candidates are GHZ states for qubits and NOON states for high-dimensional systems. A physical encoding of a GHZ states is, for instance, given by a collection of atoms all of which are localised around a point $0$ (denoted by the quantum state $|0\rangle$) and coherently localised around a point $1$ (denoted by $|1\rangle$). This is written as: $|00...0\rangle + |11...1\rangle$. In the mode notation it can also be written as $|N, 0\rangle + |0, N\rangle$ where the first cat denotes the mode corresponding to location $0$, and the second denotes the mode corresponding to location $1$. What makes these states macroscopic is that the two alternatives that are superposed can easily be discriminated (even if our measurements are imperfect \cite{sang}). This relies on $N$ being large enough (for a given size of the error in measurement). GHZ states are also fragile with respect to noise: it is sufficient to trace out one qubit to destroy the coherence in the whole state. In terms of dephasing, the rate for GHZ states is $n$ times larger than that for individual qubits, assuming they dephase independently and at the same rate -- likewise for NOON states if we think about particle loss \cite{hwang}.

It is not difficult to guess a measure of macroscopicity that will be maximized for the GHZ and NOON states, since these states achieve the best resolution in quantum metrology. Resolution is inversely proportional to the dispersion of the operator of the form $\sum_n \sigma_n$. For GHZ states this scales as $n$ (c.f. the best product states cannot do better than $\sqrt{n}$). General quantum states will have scaling of the form $n^p$, where $1/2\leq p \leq 1$. Following this logic, we can use the parameter $p$ to quantify quantum macroscopicity in the sense that states with $p$ closer to $1$ are considered more macroscopic (as proposed in \cite{shimizu}). This measure also confirms our intuition that W states, which are symmetric superpositions of $n$ qubits of which only one is in the state $1$ while the rest are $0$  (such as the state $|001\rangle + |010\rangle + |100\rangle$ for three qubits), are not macroscopic regardless of how high $n$ is. The key point is that the W state is entangled, but different states comprising the entangled state are not macroscopically distinct. For continuous variable systems an analogous measure was discussed in \cite{hwang}.

None of the experiments reviewed here, however, can be said to be creating GHZ or NOON states with large $n$.  For the sake of argument, let us consider the diffraction of an object through $n$ slits. If the transmission amplitude is the same for all slits, the state ends up being a W state. A redeeming feature could be thought to be the size of the object. In this case, we could introduce an extra parameter $m$ to quantify its complexity (say the number of atoms, or vibrational modes in the buckyball). Even for two slits we could now write the state as $|m,0\rangle + |0,m\rangle$, which is a NOON state. We could then argue that making $m$ bigger makes the system more macroscopic. However, here the system with $m$ degrees of freedom comes as a package and not as a non-interacting collection of $m$ atoms or other constituents. One could perhaps argue that in diffraction GHZ states could be realised by all $n$ objects going through $n$ slits, and simultaneously by not going through them. This, however, would be a superposition of different masses that is very hard to envisage (and, even if allowed, would last for a very short time).

Other experiments described above could be similarly analysed and shown to fall short of the GHZ ideal. This, of course, does not necessarily mean that they are not quantum macroscopic. We do not really understand what the full class of highly macroscopic quantum states ought to be (other than the GHZ and NOON states). The wide variety of systems we considered required us to assess each on its own merits and to take into account its specific parameters as well as its interaction with the environment, including the measurement apparatus. This makes comparison between the different systems challenging. But it also raises the prospect of codifying a systematic basis for comparing the different systems. Until we have a well-defined theoretical notion and measure of quantum macroscopicity, we will most likely need to rely on our more basic intuition as to what properties such quantum states should satisfy.

Interestingly, the amount of entanglement, as measured by some kind of distance to the closest separable states, is not a good indicator of macroscopicity. By this measure GHZ states contain only one bit of entanglement (independently of the number of qubits). Likewise, there are states that are highly entangled in terms of the distance measure, but fail to show a high degree of macroscopicity. Perhaps there is as yet an undiscovered trade-off between the macroscopicity of a state and its distance to the separable states~\cite{Vedral-2014}.

In closing, it is worth mentioning that the debate regarding macroscopicity is topical precisely because we are not in possession of a large scale quantum computer. A large universal quantum computer would, by definition, be able to prepare any desired quantum state deterministically (with a negligible error). In our quantum experiments, on the other hand, we usually aim at preparing a specific quantum state, such as that of a superfluid or the ground state of a micro-mechanical oscillator.  Fundamental and technology-motivated research, advances in measurement techniques and micro-fabrication over the next few years will continue to drive the discovery of ever more impressive examples of macroscopic quantum behaviour. But the fundamental question as to how far we can push the quantum limit into the macroscopic realm is likely to remain unanswered for some time, most probably until quantum computers capable of coherently manipulating thousands of qubits are finally built.

\medskip

\noindent {\bf Acknowledgments}
The authors thank Dr Wonmin Son for useful insights and comments and gratefully acknowledge financial support from the Oxford Martin Programme on Bio-Inspired Quantum Technologies, the Singapore Ministry of Education and National Research Foundation (Singapore), the EPSRC (UK), the Fell Fund (Oxford) and the Leverhulme Trust (UK). T.F. is James Martin Fellow of the Oxford Martin School, Oxford, and V.V. is Fellow of Wolfson College, Oxford.

%\begin{references}

%\end{references}
\end{document}